\documentclass{pasj00}

\begin{document}
\SetRunningHead{T. Totani et al.}{Reionization and 
the Spectrum of GRB 050904 at 
$z = 6.3$}
\Received{}
\Accepted{}

\title{Implications for the Cosmic Reionization from the Optical
Afterglow Spectrum of the Gamma-Ray Burst 050904 
at $z = 6.3$\thanks{Based on data collected at the Subaru Telescope, which
is operated by the National Astronomical Observatory of Japan.}}

\author{
Tomonori \textsc{Totani}\altaffilmark{1}, 
Nobuyuki \textsc{Kawai}\altaffilmark{2}, 
George \textsc{Kosugi}\altaffilmark{3}, 
Kentaro \textsc{Aoki}\altaffilmark{4}, 
Toru \textsc{Yamada}\altaffilmark{3}, \\
Masanori \textsc{Iye}\altaffilmark{3}, 
Kouji \textsc{Ohta}\altaffilmark{1}, 
and 
Takashi \textsc{Hattori}\altaffilmark{4}
}

\altaffiltext{1}{Department of Astronomy, Kyoto University, Sakyo-ku,
  Kyoto 606-8502} \email{totani@kusastro.kyoto-u.ac.jp}
  
\altaffiltext{2}{Department of Physics, Tokyo Institute of Technology,
  2-12-1 Ookayama, Meguro-ku, Tokyo 152-8551}

\altaffiltext{3}{National Astronomical Observatory of Japan, 2-21-1
  Osawa, Mitaka, Tokyo 181-8588 }

\altaffiltext{4}{Subaru Telescope, National Astronomical Observatory of
  Japan, Hilo, HI 96720, USA}


\maketitle

\begin{abstract}
The gamma-ray burst (GRB) 050904 at $z = 6.3$ provides the first
opportunity of probing the intergalactic medium (IGM) by GRBs at the
epoch of the reionization. Here we present a spectral modeling analysis
of the optical afterglow spectrum taken by the Subaru Telescope, aiming
to constrain the reionization history. The spectrum shows a clear
damping wing at wavelengths redward of the Lyman break, and the wing
shape can be fit either by a damped Ly$\alpha$ system with a column
density of $\log (N_{\rm HI}/{\rm cm^{-2}}) \sim 21.6$ at a redshift
close to the detected metal absorption lines ($z_{\rm metal} = 6.295$),
or by almost neutral IGM extending to a slightly higher redshift of
$z_{\rm IGM,u} \sim 6.36$. In the latter case, the difference from
$z_{\rm metal}$ may be explained by acceleration of metal absorbing
shells by the activities of the GRB or its progenitor. However, we
exclude this possibility by using the light transmission feature around
the Ly$\beta$ resonance, leading to a firm upper limit of $z_{\rm IGM,u}
\leq 6.314$. We then show an evidence that the IGM was largely ionized
already at $z=6.3$, with the best-fit neutral fraction of IGM, $x_{\rm
HI} = 0.00$, and upper limits of $x_{\rm HI} < 0.17$ and 0.60 at 68 and
95\% C.L., respectively.  This is the first direct and quantitative
upper limit on $x_{\rm HI}$ at $z > 6$. Various systematic uncertainties
are examined, but none of them appears large enough to change this
conclusion. To get further information on the reionization, it is
important to increase the sample size of $z \gtrsim 6$ GRBs, in order to
find GRBs with low column densities ($\log N_{\rm HI} \lesssim 20$)
within their host galaxies, and for statistical studies of Ly$\alpha$
line emission from host galaxies.
\end{abstract}


\section{Introduction}
\label{section:introduction}

Although more than 30 years have passed since the discovery of gamma-ray
bursts (GRBs) (Klebesadel, Strong, \& Olson 1973, see, e.g.,
M\'esz\'aros 2002; Piran 2004 for recent reviews), it was rather recent
that GRBs became widely recognized as a unique tool of cosmological
studies and exploring the early universe.  It was just before GRBs were
proven to have the cosmological origin (Metzger et al. 1997) that the
first attempt to study the cosmic star formation history by using 
GRBs was made (Totani 1997; Wijers et al. 1998). The potential use of GRBs as a
probe of the reionization history of the intergalactic medium (IGM) was
also pointed out (Miralda-Escud\'e 1998), but at that time, it was
thought that GRBs can probe the universe at most up to modest redshifts
of $z \sim $ 3 by instruments available at that time or in the near
future.  Such notion was, however, soon discarded by the discoveries of
extremely luminous GRBs like GRB 971214 (Kulkarni et al. 1998) and
990123 (Kulkarni et al. 1999), which could be detected even at redshifts
beyond 10. Then it did not take long time before astrophysicists started
to discuss GRBs as a promising lighthouse to study the extremely high
redshift universe, potentially giving cosmologically important
information including the population III star formation and the
reionization history (Lamb \& Reichart 2000; Ciardi \& Loeb 2000).

During 2000--2004, satellites such as the BeppoSAX and the HETE-2
continued to discover more and more GRBs, and there was important
progress including the establishment of the firm connection between long
duration GRBs and energetic supernovae (Stanek et al. 2003; Hjorth et
al. 2003a).  However, the distance of GRBs did not extend to very high
redshift, with the highest record of $z = 4.5$ (Andersen et
al. 2000). The launch of the Swift satellite in 2004 then allowed the
GRB community to search for fainter and more distant GRBs with improved
detection rate. The GRB 050904 was discovered by the Swift on 2005
September 4 at 01:51:44 UT (Cusumano et al. 2005), and follow-up
photometric observations of the afterglow found a strong spectral break
between optical and near-infrared (NIR) bands, indicating a very high
redshift of $z \sim 6$ (Haislip et al. 2005; Price et al. 2005;
Tagliaferri et al. 2005). This suggestion was confirmed by the
subsequent spectroscopic observation by the Subaru Telescope, which
found metal absorption lines at $z = 6.295$ and the corresponding Lyman
break and red damping wing (Kawai et al. 2005, hereafter Paper I).  This
opens a new era of GRB observations at redshifts that are close to the
cosmic reionization and comparable to those of the most distant galaxies
(Taniguchi et al. 2005) and quasars (White et al. 2003; Fan et
al. 2006).  Here we report a detailed analysis and interpretation of the
optical afterglow spectrum of GRB 050904 presented in Paper I, to derive
implications for the reionization.

The famous Gunn-Peterson (GP) test tells us that the IGM is highly
ionized at $z \lesssim 5$ (Gunn \& Peterson 1965), while the
observations of the cosmic microwave background radiation (CMB)
indicates that the universe became neutral at the recombination epoch of
$z \sim 1100$ (Spergel et al. 2003). The reionization of the IGM is
believed to have occurred during $z \sim$ 6--20 by the first stars
and/or quasars, and the precise epoch and nature of the reionization is
one of the central topics in the modern cosmology (see Loeb \& Barkana
2001; Barkana \& Loeb 2001; Miralda-Escud\'e 2003; Haiman 2004 for
recent reviews).  The dramatic increase of the optical depth of the
Ly$\alpha$ forest with increasing redshift at $z \gtrsim 5.2$
(Djorgovski et al. 2001) and the subsequent discovery of broad and black
troughs of Ly$\alpha$ absorption (the GP troughs) in the spectra of $z
\gtrsim 6$ quasars (Becker et al. 2001; White et al. 2003; Fan et
al. 2003, 2006) indicate that we are beginning to probe the epoch of
reionization (Fan et al. 2002; Cen \& McDonald 2002; but see also
Songaila \& Cowie 2002).  On the other hand, the polarization
observation of the CMB by the Wilkinson Microwave Anisotropy Probe
(WMAP) indicates a much higher redshift of reionization, $z = 17
\pm 5$ (Kogut et al. 2003). Some theorists have argued that the hydrogen
in the IGM could have been reionized twice (Wyithe \& Loeb 2003; Cen
2003).

Because the cross section of the Ly$\alpha$ resonance absorption is so
large, the light blueward of the Ly$\alpha$ wavelength at the source is
completely attenuated if the IGM neutral fraction $x_{\rm HI} \equiv
n_{\rm HI}/ n_{\rm H}$ is larger than $\sim 10^{-3}$, and hence the
Ly$\alpha$ trough of $z\sim 6$ quasars gives a constraint of only
$x_{\rm HI} \gtrsim 10^{-3}$.  The cross section becomes much smaller
for longer wavelength photons than the Ly$\alpha$ resonance, and the
spectral shape of the red damping wing of the GP trough can potentially
be used to measure $x_{\rm HI}$ more precisely (Miralda-Escud\'e
1998). However, applying this method to quasars is problematic because
of the uncertainties in the original unabsorbed quasar spectra and the
proximity effect (Bajtlik, Duncan, \& Ostriker 1988), i.e., the
ionization of surrounding IGM by strong ionizing flux from quasars (Cen
\& Haiman 2000; Madau \& Rees 2000). Therefore, though some authors
suggested $x_{\rm HI} \gtrsim 0.1$ using $z \sim 6$ quasar spectra
(Mesinger \& Haiman 2004; Wyithe \& Loeb 2004; Wyithe, Loeb, \& Carilli
2005), these estimates are generally model dependent.

The Ly$\alpha$ line emission is seriously attenuated if it is embedded
in the neutral IGM, and hence the Ly$\alpha$ line emissivity of galaxies
at $z \gtrsim$ 6 is another probe of the reionization. Therefore the
detection of Ly$\alpha$ emission from many galaxies at $z \gtrsim 6$ (Hu
et al. 2002; Kodaira et al. 2003; Taniguchi et al 2005) may indicate
that the universe had already been largely ionized at that time
(Malhotra \& Rhoads 2004; Stern et al. 2005; Haiman \& Cen 2005).
However it should not be naively interpreted as implying small $x_{\rm
HI}$, since these Ly$\alpha$ emitters (LAEs) are selected by strong
Ly$\alpha$ emission and hence they may be biased to those in ionized
bubbles created by themselves or clusters of undetected sources (Haiman
2002; Wyithe \& Loeb 2005).  On the other hand, the Lyman-break galaxies
(LBGs) selected by broad-band colors are free from the selection bias
about Ly$\alpha$ emission, but the statistical nature of Ly$\alpha$
emission from LBGs at $z \gtrsim 5$ is not yet well understood (Ando et
al. 2004; Bouwens et al. 2004; Dickinson et al. 2004; Giavalisco et
al. 2004; Stanway et al. 2004), compared with those at $z \sim 3$
(Shapley et al. 2003). Furthermore, the Ly$\alpha$ line emission from
lower-redshift starburst galaxies is often redshifted with respect to
the systemic velocity of galaxies (e.g., Pettini et al. 2000), and such
a relative redshift will lead to a higher detectability of LAEs at $z
\gtrsim 6$, indicating a possible systematic uncertainty in the
reionization study by Ly$\alpha$ emission. 

GRBs have a few advantages as a probe of the cosmic reionization,
compared with quasars or LAEs/LBGs. GRB afterglows are much brighter
than LAEs/LBGs and comparable to or even brighter than quasars if they
are observed quickly enough after the explosion. 
The Ly$\alpha$ or ultraviolet luminosity of host galaxies
is irrelevant to the detectability of GRBs, and hence GRBs can
probe less biased regions in the early universe, while quasars and
bright LAEs/LBGs are likely biased to regions of rapid structure
formation with strong clustering.  In most cases it is expected that the
IGM ionization state around GRB host galaxies had not yet been
altered by strong ionizing flux from quasars. Finally, the spectrum of
GRB afterglows has a much simpler power-law shape than complicated lines
and continuum of quasars and LAEs/LBGs, and hence model uncertainty can
greatly be reduced. Especially, a detailed fitting analysis of the
damping wing of the GP trough in an afterglow spectrum may lead to a
precise determination of $x_{\rm HI}$.

In this paper we take these advantages for the first time and derive the
first implications for the reionization from GRBs by using the spectrum
of GRB 050904. In \S \ref{section:spectrum}, we briefly describe the
overall features of the observed spectrum. Formulations of the model
fitting will be given in \S \ref{section:model}.  Before the fitting,
theoretically possible ranges of some model parameters will be defined
(\S \ref{section:z-relations}). Following the description of the fitting
procedure (\S \ref{section:procedure}), the fitting results to the
red damping wing will be given in \S \ref{section:separate-fit} and \S
\ref{section:joint-fit}. The additional constraint from the Ly$\beta$
feature is discussed in \S \ref{section:ly-beta}. We then derive a
quantitative constraint on the IGM neutral fraction, taking into account
all constraints and identified uncertainties in \S \ref{section:x_HI}. 
In \S \ref{section:discussion} we discuss the prospects of the future
GRB data as the reionization probe, based on the lessons from this first
data set; \S \ref{section:discussion-DLA} is for the neutral hydrogen
column density within host galaxies, and \S
\ref{section:discussion-lya-emission} is for the Ly$\alpha$ line
emission from host galaxies. A summary and conclusions will be presented
in \S \ref{section:summary}.

Throughout this paper, we use the WMAP values of the cosmological
parameters in the flat universe: $H_0 = 71 \ \rm km \ s^{-1} Mpc^{-1}$,
$\Omega_B = 0.044$, and $\Omega_M = 0.27$ (Spergel et al. 2003), and the
primordial helium mass fraction in the total cosmic baryon, $Y_p = 0.25$
(Kawasaki, Kohri, \& Moroi 2005).

\section{The Afterglow Spectrum of GRB 050904}
\label{section:spectrum}

\begin{figure}
  \begin{center}
    \FigureFile(160mm,160mm){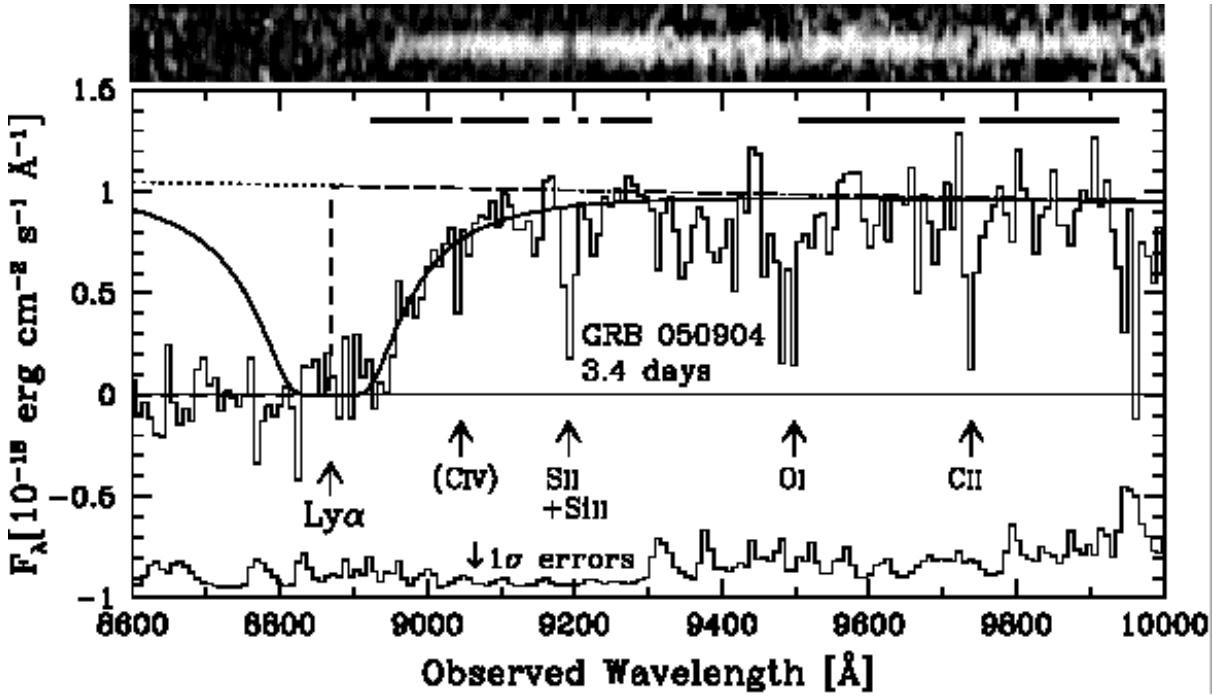}
  \end{center}
  \caption{ The afterglow spectrum of GRB 050904 taken 3.4 days after
   the burst. The spectrum is binned by 3 pixels, and 1$\sigma$ errors
   are also shown with an offset of $-1.0$. The CCD image of the
   spectrum in the corresponding wavelength range is shown at the top of
   the figure. The Ly$\alpha$ resonance and identified absorption lines
   are indicated with the redshift $z_{\rm metal} = 6.295$, except for
   the intervening C$\emissiontype{IV}$ system at $z = 4.840$. The thick
   horizontal lines in the upper right region show the wavelength ranges
   used in the spectral fitting, where identified absorption features
   are removed.  The solid curve shows the model absorption by a DLA
   with $\log N_{\rm HI} = 21.62$ and $z_{\rm DLA} = 6.295$ (the
   DLA-dominated model). The dotted line shows the original unabsorbed
   spectrum of the afterglow, with the spectral index of $\beta_0 =
   -1.25$. The dashed curve shows the model absorption by the IGM
   with $z_{\rm IGM,u} = 6.295$ and $x_{\rm HI} = 10^{-3}$, which is
   almost a vertical line at the Ly$\alpha$ resonance. The Galactic
   extinction of $E(B-V) = 0.060$ mag is taken into account in all the
   model curves. } \label{fig:spec1_v2}
\end{figure}

Figure \ref{fig:spec1_v2} presents the spectrum of GRB 050904 at
wavelengths longer than the Lyman break, taken by the FOCAS instrument
(Kashikawa et al. 2002) mounted on the Subaru Telescope (Iye et
al. 2004). Figure \ref{fig:spec2} presents the close-ups of the
Ly$\alpha$ and Ly$\beta$ resonance regions. See Paper I for the
spectrum in the full wavelength range, the details of the observation,
and basic results including the list of line detections. The pixel scale
is 2.67 \AA/pix and the wavelength resolution of the spectrum is 8.5 \AA
\ (FWHM) at $\lambda_{\rm obs} \sim 9000$\AA. The total exposure time is
4.0 hours and the mid time of the exposure was September 7, 12:05 UT
(3.4 days after the burst).

A clear damping wing is seen only redward of $\lambda_{\rm obs} \sim
8900$ \AA, corresponding to the Ly$\alpha$ wavelength
$\lambda_\alpha = c/\nu_\alpha$ = 1215.67\AA \ at the redshift of the
identified metal absorption lines, $z_{\rm metal} = 6.295$.
All the metal lines were identified at the same redshift within
the error of $\pm$ 0.002, except for an intervening absorption 
line system at $z = 4.840$. We searched any time variability of
the damping wing shape and the absorption line strength by separating
the spectrum into the first 1.5 hrs, the mid 1 hr, and the final 1.5
hrs, but statistically significant variability was not found.
 
At wavelengths redward of the Ly$\alpha$, the spectral energy
distribution (SED) has almost no feature except for the damping wing and
the metal absorption lines, as expected from the power-law SED of GRB
afterglows. There are some apparent emission-like features at
$\lambda_{\rm obs} \sim$ 8820--8920\AA, which may be Ly$\alpha$
emission. However, the data of this wavelength range is relatively noisy
and we cannot confirm these signals to be real. (See the error and the
CCD image of the spectrum shown in Fig. \ref{fig:spec2}.)  We thus set
only a conservative upper limit on the possible Ly$\alpha$ emission (\S
\ref{section:discussion-lya-emission}).

In wavelengths blueward of the Ly$\alpha$, the spectrum remarkably
resembles those of quasars at similar redshifts.  The deep, black
Ly$\alpha$ trough can be seen in $\lambda_{\rm obs} \sim$ 8500--8900 \AA
\ where the flux level is consistent with zero. On the other hand,
transmission of the afterglow light can be seen in $\sim$ 7500--8500 \AA
\ (Paper I), implying that the universe became more
transparent against the Ly$\alpha$ absorption at $z \lesssim 6$. The
transmission, however, again disappears at $\lambda_{\rm obs} \lesssim
7500$ \AA \ corresponding to the expected trough by the Ly$\beta$
absorption.

\section{Spectral Modeling}
\label{section:model}

The clear red damping wing tells us the existence of a large amount of
neutral hydrogen along the line of sight to the GRB. There are two
possibilities for the location of the neutral hydrogen. One is a damped
Ly$\alpha$ system (DLA) associated with the host galaxy, as often seen
in the optical spectra of high redshift GRB afterglows (Jensen et
al. 2001; Fynbo et al. 2001; M$\o$ller et al. 2002; Hjorth et al. 2003b;
Vreeswijk et al. 2004; Chen et al. 2005; Starling et al. 2005; Vreeswijk
et al. 2005; Watson et al. 2005b; Berger et al. 2005a). The other is the
absorption by neutral hydrogen in the IGM before the completion of the
reionization, as discussed in \S \ref{section:introduction}. Here we try
to fit these two components to the observed shape of the damping wing,
and constrain the parameters of the neutral hydrogen column density of
the DLA, $N_{\rm HI}$ (in cm$^{-2}$), and the neutral hydrogen fraction
in the IGM, $x_{\rm HI}$.

We assume the original afterglow spectrum before absorptions as $F_\nu
\propto \nu^{\beta_0}$ with the intrinsic power index of $\beta_0$.
There are a few reports of $\beta_{\rm obs}$ in the NIR bands based on
color observations: $\beta_{\rm obs} = -1.25^{+0.15}_{-0.14}$ at 10.6
hrs (Haislip et al. 2005), $\beta_{\rm obs} = -0.3 \pm 0.6$ at 0.51 days
(Price et al. 2005), and $\beta_{\rm obs} = -1.25 \pm 0.25$ at 1.155
days (Tagliaferri et al. 2005).  The NIR colors do not show any
indication of extinction at the host galaxy (Tagliaferri et
al. 2005). The X-ray data taken by the Swift shows a large soft X-ray
absorption corresponding to $\log N_{\rm HI} = 22.86$ in the first
$\sim$250 s from the trigger, but it disappears from the data after
$\sim$500 s with an upper bound of $\log N_{\rm HI} < 21.57$ (Bo\"er et
al. 2005), corresponding to a visual extinction of $A_V < 2.1$ by the
standard correlation for the solar abundance (Predehl \& Schmitt
1995)\footnote{The column densities derived by Bo\"er et al. (2005) are
also assuming the solar abundance (M. Bo\"er, a private
communication). Since the soft X-ray absorption is dominated by metals,
this conversion is not sensitive to a hydrogen-to-metal conversion
(e.g., Watson et al. 2005b).  On the other hand, the detection of X-ray
absorption for GRB 050904 may not be significant (see Watson et
al. 2005a).}.  Later in this paper we will find that $\log N_{\rm HI}
\lesssim 21.6$ from the fitting to the observed damping wing shape, and
metallicity of $Z \sim 0.1 (N_{\rm HI}/10^{21.6})^{-1} Z_\odot $ from
the S$\emissiontype{II}$ column density. By using the relation inferred
for the Small Magellanic Cloud (SMC) having a similar metallicity
(Bouchet et al. 1985), the H$\emissiontype{I}$ column density translates
to $A_V \lesssim 0.3$. In fact, the observed
reddening of optical GRB afterglows is generally even smaller than those
expected from the corresponding X-ray absorption or column densities of
H$\emissiontype{I}$ or metals (Galama \& Wijers 2001; Savaglio \& Fall
2004; Watson et al. 2005b; De Pasquale et al. 2005), possibly due to
sublimation of dust grains by strong GRB radiation (Waxman \& Draine
2000; Fruchter et al. 2001). Therefore we assume $A_V = 0$ at the host
galaxy in our baseline model.

According to the standard synchrotron shock model of GRB afterglows
(Sari, Piran, \& Narayan 1998), the observed index is rather soft
corresponding to the electron spectral index of $p = 2\beta_0 - 1 \sim
-3.5$, indicating that the frequency of the cooling break in the
afterglow SED had already passed through the NIR bands to longer
wavebands. Therefore we do not expect a spectral change between the
epochs of the photometric observations and our spectroscopic
observation.  We then take $\beta_0 = -1.25$ in our baseline model.
However, both the change of $\beta_0$ and the effect of possible
reddening by dust in the host galaxy will be discussed later.

The optical depth of the DLA is calculated in the standard manner:
$\tau_{\rm DLA}(\lambda_{\rm obs}) = N_{\rm HI} \ \sigma_\alpha[\nu_{\rm
obs} (1+z_{\rm DLA})]$, where $\nu_{\rm obs} = c/\lambda_{\rm obs}$ is
the observed frequency and $z_{\rm DLA}$ is the redshift of the DLA.
The damping wing shape is much wider than the Doppler broadening by a
reasonable velocity dispersion ($\sim$ 3 \AA \ for 100 km/s), and hence
we ignore it. We use the exact formula of the Ly$\alpha$
absorption cross section for a restframe frequency $\nu$:
\begin{eqnarray}
\sigma_\alpha(\nu) = \frac{3 \lambda_\alpha^2 f_\alpha \Lambda_{\rm cl,
\alpha}}{8 \pi} 
\frac{ \Lambda_\alpha (\nu / \nu_\alpha)^4}
{4 \pi^2 (\nu - \nu_\alpha)^2 + \Lambda_\alpha^2 (\nu/\nu_\alpha)^6/4} \ ,
\end{eqnarray}
where $f_{\alpha} = 0.4162$ and $\Lambda_\alpha = 3 (g_u/g_l)^{-1}
f_\alpha \Lambda_{\rm cl, \alpha}$ are the absorption oscillator
strength and the damping constant of the Ly$\alpha$ resonance (Peebles
1993; Madau \& Rees 2000), respectively, and $g_u$ and $g_l$
($g_u/g_l = 3$ for Ly$\alpha$) are the statistical weights for the upper
and lower levels, respectively (Cox 2000). Here, the classical
damping constant is $\Lambda_{\rm cl, \alpha} = (8 \pi^2 e^2)/(3 m_e c
\lambda_\alpha^2) = 1.503 \times 10^9 \ \rm s^{-1}$, where $e$ and $m_e$
are the electron charge and mass, respectively.

For the optical depth of the red damping wing by IGM absorption,
we use the formula given by Miralda-Escud\'e (1998):
\begin{eqnarray}
\tau_{\rm IGM}(\lambda_{\rm obs}) 
&=& \frac{x_{\rm HI} \ R_\alpha \ \tau_{\rm GP}(z_{\rm host})}{\pi} \
\left( \frac{1 + z_{\rm obs}}{1+z_{\rm host}} \right)^{3/2}
\nonumber \\ 
&\times& \left[ I\left( \frac{1+z_{\rm IGM,u}}{1+z_{\rm obs}} \right) -
I\left( \frac{1+z_{\rm IGM, l}}{1+z_{\rm obs}} \right) 
\right] \ ,
\label{eq:tau-igm}    
\end{eqnarray}
where $(1+z_{\rm obs}) \equiv \lambda_{\rm obs}/\lambda_\alpha$,
$R_\alpha \equiv \Lambda_\alpha \lambda_\alpha / (4\pi c) = 2.02 \times
10^{-8}$ and $z_{\rm host}$ is the redshift of the GRB host galaxy.  In
this formula, the IGM is assumed to be uniformly distributed in a
redshift range from $z_{\rm IGM,l}$ to $z_{\rm IGM,u}$. Since the
damping wing is by definition optically thin, the clumpiness of the IGM
does not have significant effect. We can get a reasonable constraint on
the mass-weighted optical detph by assuming a uniform distribution,
which is another advantage of using the damping wing.
The function
$I(x)$ is given by:
\begin{eqnarray}
I(x) &=& \frac{x^{9/2}}{1-x} + \frac{9}{7} x^{7/2}
+ \frac{9}{5} x^{5/2} + 3 x^{3/2} + 9 x^{1/2} \nonumber \\
&-& \frac{9}{2} \ln
\frac{1+x^{1/2}}{1 - x^{1/2}} \ ,
\end{eqnarray}
which is almost always a sufficiently good approximation under the
condition of $(z_{\rm obs} - z_{\rm IGM,u}) \gg R_\alpha (1+z_{\rm obs})$. 
The Gunn-Peterson optical depth is given as:
\begin{eqnarray}
\tau_{\rm GP}(z) &=& 
\frac{3 f_\alpha \Lambda_{\rm cl, \alpha}
 \lambda_\alpha^3 \ \rho_c \Omega_B (1-Y_p)}
{8 \pi m_p H_0 \Omega_M^{1/2}} (1+z)^{3/2} \\
&=& 3.88 \times 10^5 \left( \frac{1+z}{7} \right)^{3/2} \ ,
\end{eqnarray}
where $m_p$ is the proton mass and $\rho_c \equiv 3 H_0^2 / (8\pi G)$ is
the standard critical density of the universe at the present time. The
$1 \sigma$ uncertainty of this numerical factor coming from the
estimated errors of the cosmological parameters (Spergel et al. 2003;
Kawasaki, Kohri, \& Moroi 2005) is less than 8\%.  It should be noted
that, even though the dependence on $z_{\rm host}$ appears explicitly in
eq. ($\ref{eq:tau-igm}$) by the normalization of $\tau_{\rm GP}$ at the
host redshift, it actually cancels out with the factor $(1 + z_{\rm
host})^{-3/2}$ in eq. ($\ref{eq:tau-igm}$) and hence $\tau_{\rm IGM}$
depends only on $z_{\rm IGM,l}$ and $z_{\rm IGM,u}$. Therefore $z_{\rm
host}$ is not included in our fitting parameters.  We set $z_{\rm IGM,l}
= 6$, and dependence on this parameter will be discussed later. The
possible range of $z_{\rm IGM, u}$ will be discussed in detail in \S
\ref{section:z-relations}.

Although the neutral fraction $x_{\rm HI}$ cannot exceed the unity,
what is relevant for the IGM absorption is the absolute density of
neutral hydrogen rather than the neutral fraction, 
and the total cosmic hydrogen density depends on
the cosmological parameters. Therefore, we allow $x_{\rm HI} > 1$ 
in this paper, as a parameter meaning 
the neutral hydrogen density normalized
by the total hydrogen density estimated from the standard values of the
cosmological parameters.

\section{Possible Ranges of $z_{\rm DLA}$ and $z_{\rm IGM, u}$}
\label{section:z-relations}

Now we have the primary fitting parameters of $N_{\rm HI}$, $z_{\rm
DLA}$, $x_{\rm HI}$, and $z_{\rm IGM, u}$.  Though $z_{\rm DLA}$ and
$z_{\rm IGM,u}$ can be treated as free parameters in the fitting, metal
absorption lines at $z_{\rm metal} = 6.295$ have been clearly detected
in the spectrum of GRB 050904. Therefore theoretical expectation and
possible ranges of $z_{\rm DLA}$ and $z_{\rm IGM, u}$ with respect to
$z_{\rm metal}$ should be discussed.

The most clearly detected feature is S\emissiontype{II} $\lambda$1259.5
+ Si\emissiontype{II} $\lambda$1260.4, and S\emissiontype{II}
$\lambda$1253.8 and Si\emissiontype{II}* $\lambda$1264.7 are also
detected beside the 1260 \AA \ feature. The profile fitting and curve of
growth analysis revealed that these lines are not
heavily saturated and only marginally resolved, giving column densities
of $\log N_{\rm SII} = 15.60^{+0.14}_{-0.17}$ and $\log N_{\rm SiII} =
14.29^{+0.57}_{-0.39}$ and a velocity dispersion of $\lesssim$ a few
hundreds km/s (Paper I). The fine-structure silicon,
Si$\emissiontype{II}$*, is often detected in GRB afterglow spectra
(Vreeswijk et al. 2004; Chen et al. 2005; Berger et al. 2005a, b) and
indicates a high density environment, while it has never been clearly
detected from DLAs in quasar (QSO) spectra. This suggests that the metal
absorption system is a high density gas at the vicinity of the GRB
explosion site (such as star forming regions, molecular clouds, or
shells ejected by the GRB progenitor's activity), rather than by absorption
on a galactic scale of the host galaxy or by an intervening system.

\subsection{$z_{\rm DLA}$ versus $z_{\rm metal}$}
\label{section:DLA-vs-metal}

A clearly reasonable assumption is $z_{\rm DLA} = z_{\rm metal}$,
because no metal absorption lines were identified at redshifts different
from 6.295, except for the intervening system with a very different
redshift, $z = 4.840$. Later we will find that $\log N_{\rm HI} \sim
21.6$ is necessary\footnote{This column density is slightly different
from $\log N_{\rm HI} = 21.3$ adopted in Paper I, which is the best-fit
DLA model when $z_{\rm DLA}$ is treated as a free parameter (see \S
\ref{section:separate-fit}). The following metallicity estimates are
correspondingly different from those in Paper I, but the difference of
this magnitude does not affect the arguments in this paper.}  to explain
the observed damping wing shape by the DLA absorption with $z_{\rm DLA}
= z_{\rm metal}$.  This column density is typical for DLAs found in GRB
afterglows (Vreeswijk et al. 2004), and is also similar to those of
molecular clouds in the Galaxy (Larson 1981; Solomon et al. 1987). This
is consistent with the high density inferred from the
Si$\emissiontype{II}$* detection. Adopting this hydrogen column density,
we get the metallicity estimates as [S/H] = $-1.3$ and [Si/H] = $-2.9$,
using the solar abundance of Grevesse \& Sauval (1998). Here we made no
correction for ionization, assuming that the column densities of the
singly ionized ions are equal to the total column densities, which is a
reasonable assumption for silicon and sulphur in regions of large HI
absorption like DLAs (Viegas 1995; Vladilo et al. 2001). Sulphur is not
depleted to dust grains and the metallicity inferred from [S/H] is
reasonable compared with those found in other GRB-DLAs.  (See Vreeswijk
et al. 2004 for a compilation of metallicities of GRB-DLAs and
QSO-DLAs.) The large depletion of silicon compared with sulphur is not
surprising, since large dust depletions compared with QSO-DLAs have been
observed in a few GRB afterglows (Savaglio, Fall, \& Fiore 2003).
Therefore the observed column densities and metallicities are reasonable
in the case of the physical association between the DLA and absorption
lines.

On the other hand, it seems rather unlikely that the DLA and the metal
lines have different redshifts, again from the metallicity argument. In
this case, no detectable lines associated with the DLA set $2 \sigma$
upper limits on metallicities as [S/H] $\leq -1.5$ and [Si/H] $\leq
-3.1$, while the metallicity of the region associated with the metal
lines would become larger than those estimated above. Hence there must
be a large metallicity difference between the DLA and the metal
absorption system. Such metallicity inhomogeniety seems rather unlikely
if the DLA is also physically associated with the high density region in
the vicinity of the GRB. On the other hand, if the DLA is a system on
the galactic scale like QSO-DLAs in the host galaxy or in any
intervening galaxies, we do not expect heavy depletion of Si (Savaglio
et al. 2003), and hence the upper limit on [Si/H] should not be far from the
total metallicity. However, such a low metallicity is not observed in
any QSO-DLAs at $z \sim 1$--5, though there is no measurements of
QSO-DLA metallicity at $z > 6$ (Prochaska et al. 2003).  Based on these
considerations, we set $z_{\rm DLA} = z_{\rm metal}$ in our baseline
model.

However, we conservatively allow the case of $z_{\rm DLA} = z_{\rm host}
\neq z_{\rm metal}$, supposing that the absorbing neutral hydrogen is
distributed on the scale of the host galaxy.  Though a naive expectation
is $z_{\rm host} = z_{\rm metal}$, we expect a systemic velocity
difference between them up to a few hundreds km/s depending on the host
galaxy's circular velocity ($|z_{\rm host} - z_{\rm metal}| \lesssim
0.005$). Furthermore, absorption lines blueshifted up to $\sim$ 3,000
km/s with respect to the restframe of the host galaxy were observed in
the afterglow spectrum of GRB 021004, which may have been accelerated by
activities of the GRB or its progenitor (M$\o$ller et al. 2002; Schaefer
et al. 2003; Mirabal et al. 2003). Similar features were found also for
GRB 020813 (Barth et al. 2003) and GRB 030226 (Klose et al. 2004).  The
blueshifted metal lines in the GRB 021004 spectrum were highly ionized
(C\emissiontype{IV} and Si\emissiontype{IV}), but Ly$\alpha$ absorption
was observed as well at the same relative velocity. This indicates that
the accelerated absorber is a mixture of neutral and highly ionized gas,
and it is not unreasonable if blueshifted Si\emissiontype{II} is
detected in some other GRBs. 

There are, however, two arguments disfavoring the blueshift of $\sim
3,000$ km/s. First, it seems rather unlikely that only metals in such a
high velocity shell were detected without any detection at other
velocities or the burst restframe.  Second, such high velocity shells
are expected only within $\sim$ 1 pc from the burster (Schaefer et
al. 2003; Mirabal et al. 2003), which might be inconsistent with the
large amount of the silicon depletion, because strong GRB radiation
would destruct dust grains within $\sim 1$ pc of a burster (Waxman \&
Draine 2000; Fruchter et al. 2001; Draine \& Hao 2002). However, a
detailed modeling is required to examine these possibilities more
quantitatively, and here we allow a blueshift of $z_{\rm metal}$ up to
$(z_{\rm host} - z_{\rm metal}) \sim +0.07$. Therefore, the possible
range of $z_{\rm DLA}$ becomes $-0.005 \lesssim (z_{\rm DLA} - z_{\rm
metal}) \lesssim +0.07$.

\subsection{$z_{\rm IGM,u}$ versus $z_{\rm metal}$}
\label{section:z_IGMu_z_metal}

We should set $z_{\rm IGM,u} = z_{\rm host}$ if the GRB occurred in a
galaxy embedded in the IGM that had not been ionized by the emission
from the galaxy. Barkana \& Loeb (2004) estimated that typical $z \sim
7$ GRBs occur in galaxies whose dark halo mass is $M_h \sim 4 \times
10^8 M_\odot$, based on the standard structure formation theory.  They
found that the proper (i.e., not comoving) radius of the ionized bubble
created by stellar radiation from such galaxies would be $\lesssim
0.1$ Mpc, corresponding to a wavelength shift of $\Delta \lambda_{\rm
obs} \sim 3$\AA \ or $(z_{\rm IGM,u} - z_{\rm host}) \sim -0.0025$.
Instead, infall of IGM into the host galaxy may result in a relative
redshift of the IGM absorption, with a wavelength shift of $\Delta
\lambda_{\rm obs} \lesssim 1$\AA \ for a typical GRB host galaxy
(Barkana \& Loeb 2004). Since these are sufficiently small compared with
the wavelength range of the damping wing, we set $z_{\rm IGM,u} = z_{\rm
host}$ in the baseline model.  On the other hand, Haiman (2002)
estimated the proper size of ionized bubbles around LAEs detected in
deep surveys as $\sim 0.8$ Mpc ($z_{\rm IGM,u} - z_{\rm host} \sim
-0.02$), assuming a star formation rate (SFR) of $\sim 10 M_\odot$/yr
(as typically estimated for LAEs from the Ly$\alpha$ emission
luminosity) and stellar age of $10^8$ yr. This means a stellar mass
greater than $M_* \sim 10^9 M_\odot$, which is considerably larger than
that estimated by Barkana \& Loeb (2004). Hence we consider that such a
large ionized bubble is rather unlikely for {\it typical} GRBs at $z
\sim 7$.  It should be noted that the LAEs found in deep surveys are not
necessarily typical galaxies or a tracer of typical star formation at $z
\sim 7$.

Based on these considerations combined with the possible difference
of $z_{\rm host}$ from $z_{\rm metal}$ discussed in 
\S\ref{section:DLA-vs-metal}, we define the possible range of $z_{\rm IGM,u}$
as $-0.02 \lesssim 
(z_{\rm IGM,u} - z_{\rm metal}) \lesssim +0.07$, 
with a canonical value of $z_{\rm IGM,u} = z_{\rm metal}$. 

\section{Fitting to the Red Damping Wing}
\label{section:fitting}

\subsection{Fitting Procedures}
\label{section:procedure}

In this section we will fit the models described above to the observed
shape of the red damping wing. The fit is performed in a wavelength
range of $\lambda_{\rm obs} = $
8925--9938\AA. \ The lower bound corresponds to the wavelength
where the wing drops sharply to the flux zero level. The wavelengths
shorter than this are excluded since the noise by atmospheric lines is
relatively high and there may be Ly$\alpha$ emission from the host
galaxy (see discussion in \S \ref{section:discussion-lya-emission}).  We
then excluded the following wavelength regions of known absorption
lines: 9033--9046\AA \ (likely to be C\emissiontype{IV} in an
intervening system at $z = 4.840$), 9137--9158\AA \ (S\emissiontype{II}
$\lambda$1253.8), 9180--9206\AA \ (S\emissiontype{II} $\lambda$1259.5 +
Si\emissiontype{II} $\lambda$1260.4), 9218--9238\AA \
(Si\emissiontype{II}* $\lambda$1264.7), 9490--9505\AA \
(O\emissiontype{I} $\lambda$1302.1), and 9728--9750\AA \
(C\emissiontype{II} $\lambda$1334.5).  In addition, we excluded a range
of 9305--9490\AA, since there are some absorption features that are
probably atmospheric. There are 268 data points (pixels) in the final
wavelength ranges used in the $\chi^2$ analyses, which 
are indicated by thick horizontal lines in Fig. \ref{fig:spec1_v2}.

\begin{figure}
  \begin{center}
    \FigureFile(90mm,90mm){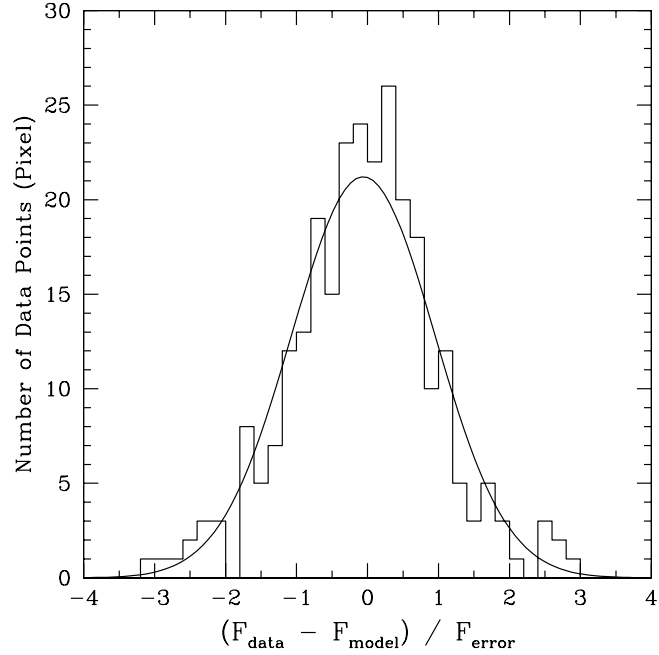}
  \end{center}
  \caption{The histogram for the deviation of the observed flux in a
  pixel from a model spectrum, normalized by the 1$\sigma$ error of the
  data.  The reference model is with $\log N_{\rm HI} = 21.62$, $z_{\rm
  DLA} = 6.295$, and $x_{\rm HI} = 0$ (the DLA-dominated model). There
  are 268 data points (corresponding to the number of pixels) in the
  wavelength range used in the fitting analysis. The solid curve is the
  fit by the Gaussian distribution with $\sigma = 1$.  }
  \label{fig:chi2_hist}
\end{figure}

Then the model spectrum is fit to the data, with normalization always
chosen to minimize the $\chi^2$ value. In addition to the DLA and IGM
absorptions, the Galactic extinction by $E(B-V) = 0.060$ mag (Schlegel,
Finkbeiner, \& Davis 1998) is also taken into account in the model
spectrum. The spectral resolution is sufficiently narrower than the
observed damping wing shape, and hence we ignore the smoothing effect by
it.  Figure \ref{fig:chi2_hist} shows the histogram of the deviation of
the data flux in a pixel from the model flux, $(F_{\rm data} - F_{\rm
model}) / F_{\rm error}$, for one of the acceptable fits with $z_{\rm
DLA}$ = 6.295, $\log N_{\rm HI} = 21.62$, and $x_{\rm HI} = 0$ ($\chi^2
= 277.29$).  The distribution can well be fit by the Gaussian
distribution, indicating that the error is well controlled and we can
derive a reasonable statistical constraint on the model parameters by
the standard $\chi^2$ analyses.

\subsection{Separate Fittings by DLA and IGM Absorption}
\label{section:separate-fit}

\begin{figure}
  \begin{center}
    \FigureFile(90mm,90mm){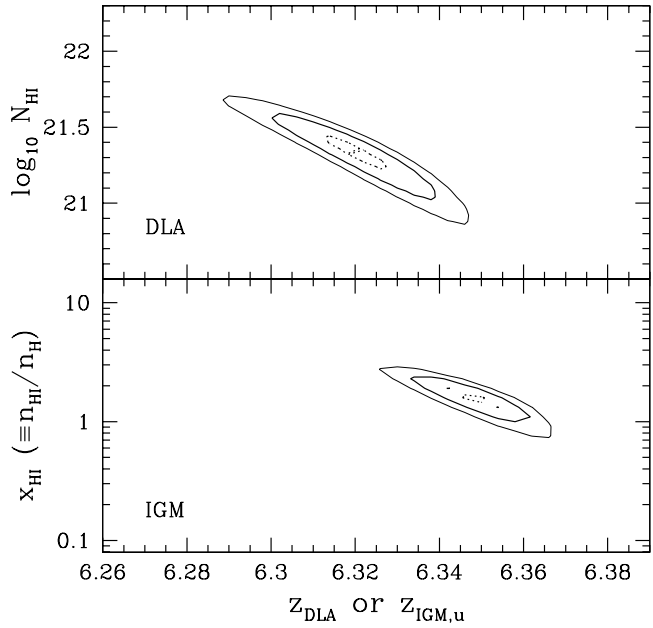}
  \end{center}
  \caption{The arrowed regions for the separate fits of the DLA (upper
   panel) and IGM (lower panel) absorption models to the red damping
   wing. The horizontal axis is the DLA redshift ($z_{\rm DLA}$) for the
   DLA model but is the upper redshift bound of the extension of neutral
   hydrogens in the IGM ($z_{\rm IGM,u}$) for the IGM model. The
   contours show the levels of $\Delta \chi^2 = 2.30, 9.21$, and $18.4$
   from the minimum $\chi^2$, corresponding to 68.3 (dotted), 99 (thick
   solid), and 99.99\% (thin solid) confidence levels for two degrees of
   freedom. See text (\S \ref{section:model}) for the reason why the
   apparently unphysical region of $x_{\rm HI} > 1$ is also shown. The
   redshift of metal absorption system and its error are indicated by
   the thick and thin vertical lines, respectively. }
   \label{fig:separate-contour}
\end{figure}

As the first simple analysis, we try to fit the two components
separately to the data. Figure \ref{fig:separate-contour} shows the
confidence region of $N_{\rm HI}$-$z_{\rm DLA}$ for the DLA fit and
$x_{\rm HI}$-$z_{\rm IGM, u}$ for the IGM fit. The best fits are $(\log
N_{\rm HI}, \ z_{\rm DLA}) = (21.34, \ 6.320)$ with $\chi^2 = 265.54$ and
$(x_{\rm HI}, \ z_{\rm IGM,u}) = (1.58, \ 6.348)$ with $\chi^2 = 270.79$,
respectively. 

The best-fit value of $z_{\rm DLA}$ is different from $z_{\rm metal}$ by
$\sim +0.025$, which is considerably larger than that expected from the
circular velocity of the host galaxy, indicating a possibility that the
metal absorption system was accelerated by the GRB or its
progenitor. However, $z_{\rm DLA} = z_{\rm metal} = 6.295$ is also
marginally consistent within the confidence region of 99.7\% C.L.,
and taking into account the arguments given in \S\ref{section:DLA-vs-metal},
we keep this case as the baseline model.

Interestingly, almost neural IGM with $x_{\rm HI} \sim 1$ extending to
$z_{\rm IGM,u} \sim 6.36$ gives a marginally acceptable fit for the red
damping wing, indicating a possibility that we may have observed for the
first time the signature of the completely neutral IGM before the
reionization.  The redshift difference of $(z_{\rm IGM,u} - z_{\rm
metal}) \sim +0.065$ is within the theoretically possible range
discussed in \S\ref{section:z_IGMu_z_metal}.

We show the model spectra for the two representative cases: ``the
DLA-dominated model'' with $z_{\rm DLA} = 6.295$ and $(\log N_{\rm HI},
\ z_{\rm IGM,u}, \ x_{\rm HI})$ = (21.62, 6.295, 0) in
Fig. \ref{fig:spec1_v2} and ``the IGM-dominated model'' with ($-\infty$,
6.36, 1.0) in Fig. \ref{fig:spec2}. Although the red damping wing shape
can be reproduced even with $x_{\rm HI} = 0$ in the DLA-dominated model,
the IGM absorption with $x_{\rm HI} \gtrsim 10^{-3}$ and $z_{\rm IGM,u}
\gtrsim 6.26$ are required for the absorption of photons blueward of the
Ly$\alpha$ resonance.

\subsection{DLA-IGM Joint Fitting}
\label{section:joint-fit}

\begin{figure}
  \begin{center}
    \FigureFile(80mm,80mm){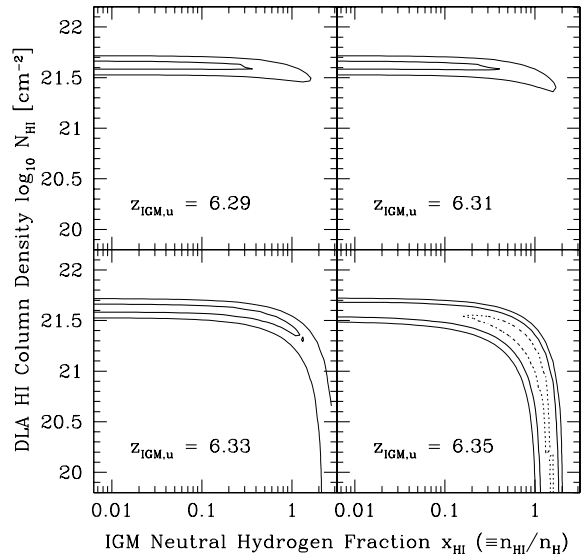}
  \end{center}
  \caption{The arrowed regions for the joint fit of the DLA and IGM
   absorption models. The four panels are for different values of
   $z_{\rm IGM,u}$ as indicated in the panels. The DLA redshift is fixed to 
   $z_{\rm DLA} = z_{\rm metal} = 6.295$. 
   The contours show the levels of $\Delta \chi^2 =
   3.53, 11.3$, and $21.1$ from the minimum $\chi^2$, corresponding to
   95 (dotted), 99 (thick solid), and 99.99\% (thin solid)
   confidence levels for three degrees of freedom.
   See text (\S \ref{section:model}) for the reason
  why the apparently unphysical region of $x_{\rm HI} > 1$ is shown.
}
  \label{fig:joint-contour}
\end{figure}

Next we try a joint fit including both the DLA and IGM absorptions.
Here we treat $z_{\rm IGM,u}$ as a free parameter but assume $z_{\rm
DLA} = z_{\rm metal} = 6.295$ to reduce the number of free papameters.
Figure \ref{fig:joint-contour} shows the allowed region for $N_{\rm HI}$
and $x_{\rm HI}$, for several values of $z_{\rm IGM,u}$.  The best-fit
is obtained at $(z_{\rm IGM,u}, \ \log N_{\rm HI}, \ x_{\rm HI}) =
(6.36, \ 21.59, \ 0.016)$ with $\chi^2 = 269.70$. Constraining $x_{\rm
HI} \geq 10^{-3}$ as inferred from the GP trough in quasar spectra at $z
\gtrsim 6$, we obtain an upper limit of $z_{\rm IGM, u} \leq 6.371$ at
99\% C.L., otherwise the IGM absorption will erase the light
transmission at $\lambda_{\rm obs} \gtrsim 8950$ \AA.

As expected from the single component fits, the damping wing can be
explained mostly by the DLA absorption at $z_{\rm DLA} = 6.295$ with
negligible IGM absorption, while it can also be fit dominantly by the
IGM absorption with $x_{\rm HI} \sim 1$ and $z_{\rm IGM,u} \sim 6.36$.
These results tell us that there is an unfavorable degeneracy between
absorptions by a DLA and IGM (Miralda-Escud\'e 1998; Barkana \& Loeb
2004).  However, in the next section we show that this degeneracy can be
broken by utilizing the features around the Ly$\beta$ resonance.

\subsection{Constraints from the Ly$\beta$ Feature}
\label{section:ly-beta}

Here we apply the same absorption models of the DLA and IGM, but for the
Ly$\beta$ transition at the restframe wavelength of $\lambda_\beta =
1025.72$ \AA. We use the same formulae given in \S \ref{section:model},
but replace $\lambda_\alpha$ and $f_\alpha$ by the corresponding values
for Ly$\beta$. The Ly$\beta$ absorption oscillator strength is $f_\beta
= 0.07910$ and $g_u/g_l =3$ is the same for all the Lyman series
transitions by the selection rules (Cox 2000; Morton 1991). The
frequency-integrated cross section $\int \sigma (\nu) d\nu$ is reduced
by $f_{\beta}/ f_\alpha$ = 0.190, and the width of the damping wing in
the wavelength space ($\propto \Lambda_\beta \lambda_\beta^2$) is
reduced by the same factor. The optical depth of the red damping wing 
($\tau_{\rm IGM}$) is reduced by a factor of $(f_{\beta}/f_{\alpha})^2
= 0.0361$ for the same $z_{\rm obs}$ ($= \lambda_{\rm obs}/\lambda_{\beta}
-1$ for Ly$\beta$). 

\begin{figure}
  \begin{center}
    \FigureFile(80mm,80mm){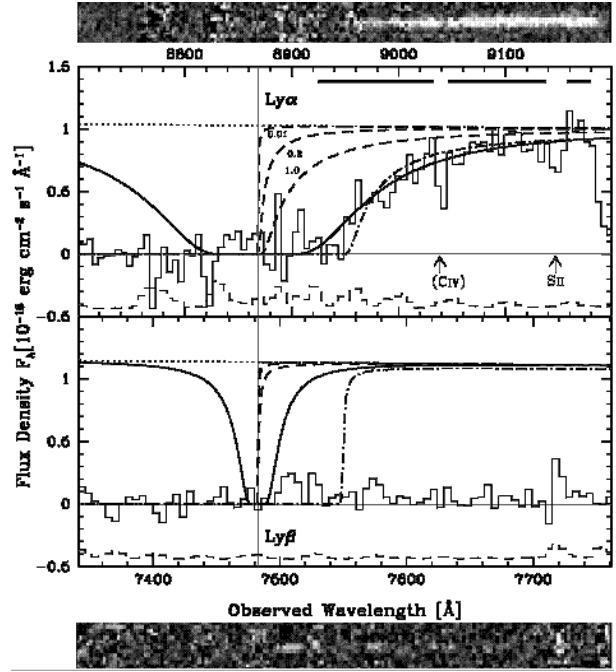}
  \end{center}
  \caption{ The same as Fig. \ref{fig:spec1_v2}, but the close-up of the
  Ly$\alpha$ (upper panel) and Ly$\beta$ (lower panel) regions. The
  spectrum is binned by 2 pixels, and 1$\sigma$ errors are shown by
  dashed lines with an offset of $-0.5$. The wavelength scales of the
  two panels are chosen so that the Ly$\alpha$ and Ly$\beta$ wavelengths
  at a fixed redshift coincide with each other on the horizontal scale
  (i.e., $\lambda_{\rm top}/\lambda_{\rm low} =
  \lambda_\alpha/\lambda_\beta$). The vertical solid line marks the
  Ly$\alpha$ and Ly$\beta$ resonances at $z = 6.295$. The CCD images of
  the spectrum in the corresponding wavelength ranges are shown at the
  top and bottom of the figure.  The solid and dotted curves are the
  same with those in Fig.  \ref{fig:spec1_v2}. The three dashed curves
  show the model absorption by the IGM ($z_{\rm IGM,u} = 6.295$), with
  $x_{\rm HI} = $ 0.01, 0.2, and 1.0, as indicated. The dot-dashed curve
  is the model absorption by the IGM but with $z_{\rm IGM,u} = 6.36$ and
  $x_{\rm HI} = 1.0$ (the IGM-dominated model).}  \label{fig:spec2}
\end{figure}

The predictions for the Ly$\beta$ absorption feature are presented
correspondingly to Ly$\alpha$ in Fig. \ref{fig:spec2}. The feature of
the DLA-dominated model is in good agreement with the data, while the
emission (or, residual of absorption) feature around 7500--7520\AA \ is
clearly inconsistent with the IGM-dominated model, in which the
transmission is completely attenuated at wavelengths blueward of
$\lambda_\beta (1+z_{\rm IGM,u}) \sim 7550$\AA.  By visual inspection of
the original CCD image of the spectrum (shown in Fig. \ref{fig:spec2}),
we confirmed that the feature of 7500--7520\AA \ is the real
signal. Therefore the Ly$\beta$ feature safely excludes the
IGM-dominated model. By requiring that the IGM Ly$\beta$ absorption does
not erase the feature at 7500--7520\AA, we set a firm upper limit of
$z_{\rm IGM,u} \leq 6.314$, and within this constraint, the IGM
absorption has only small contribution to the observed red damping wing
even if $x_{\rm HI} = 1$ (see dashed curves of Fig. \ref{fig:spec2}). 
However, this contribution is not negligible,
and in fact we can derive some constraint on $x_{\rm HI}$, as will be
shown below.

\section{Implications for the IGM Neutral Fraction around
GRB 050904}
\label{section:x_HI}

After the long analyses presented above, we finally arrived at the most
likely interpretation of the red damping wing: it is mainly contributed
by the DLA associated with the metal absorption lines at $z_{\rm
metal} = 6.295$, and the contribution from neutral IGM is not necessary
at all. However, as shown in Fig. \ref{fig:spec2}, if the IGM is
completely neutral, the IGM absorption with $z_{\rm IGM,u} \sim 6.295$ has
small but non-negligible contribution to the observed damping wing
shape. Then the next question is which value of $x_{\rm HI}$ is favored,
under the constraints of $z_{\rm IGM,u} = z_{\rm DLA} =
6.295$ but $N_{\rm HI}$ is treated as an unknown free parameter
and always chosen to minimize $\chi^2$ for each value of $x_{\rm HI}$.

For parameters other than $x_{\rm HI}$ and $N_{\rm HI}$, we take $z_{\rm
IGM,u} = z_{\rm DLA} = 6.295$, $z_{\rm IGM,l} = 6$, $\beta_0 = -1.25$,
and $A_V = 0$ as the baseline model parameters.  We then find that
$\chi^2$ becomes minimum at $x_{\rm HI} = 0$ with $\log N_{\rm HI} =
21.62$, and it continuously increases with $x_{\rm HI}$, changing the
best-fit value of $\log N_{\rm HI}$ to 21.54 at $x_{\rm HI} = 1$. [The
preference to smaller $x_{\rm HI}$ for $z_{\rm IGM,u} \lesssim 6.32$ was
also found in the results of the DLA-IGM joint fit (see \S
\ref{section:joint-fit} and Fig. \ref{fig:joint-contour}).]  Then
formally we can estimate the confidence region of $x_{\rm HI}$ by
requiring that $\Delta \chi^2 (x_{\rm HI}) \equiv \chi^2(x_{\rm HI}) -
\chi^2_{\min}$ obeys the chi-square distribution with one degree of
freedom (Lampton, Margon, \& Bowyer 1976; Press et al. 1992). Table
\ref{table:chi2} presents the $\chi^2$ values, the exclusion confidence
level of the $x_{\rm HI} = 1$ model corresponding to $\Delta \chi^2
(1)$, and upper limits on $x_{\rm HI}$ with 68 ($1\sigma$), 95, and 99
\% C.L. We find the best fit value of
$x_{\rm HI} = 0.00$ in all models and upper limits of
$x_{\rm HI} < $0.17 and 0.60 
(68 and 95\% C.L., respectively) for the baseline model.

In order to show why the $x_{\rm HI} = 1$ model gives a worse fit, we
plot in Fig. \ref{fig:spec-residual} the model spectra and the observed
data in the form of the residual from the best-fit model of $x_{\rm
HI} = 0$ (the DLA-dominated model). The worse fit of the $x_{\rm HI} =
1$ model comes from the combination of the two effects: the change of
the damping wing shape in $\lambda_{\rm obs} = $ 8900--9200 \AA \ is not
favored from the data, and the model predicts higher flux than the
$x_{\rm HI} = 0$ model at the longest wavelength range of $\lambda_{\rm
obs} \gtrsim 9500$\AA, while the data favor even lower flux than that of
the $x_{\rm HI} = 0$ model.

Before we conclude that the case of $x_{\rm HI} = 1$ is not favored,
however, any systematic uncertainties must be explored. The
results of this uncertainty check will be given below. Table
\ref{table:chi2} shows the fit results changing one or two
of the model parameters while the other parameters are kept to the values
of the baseline model. (The DLA column density $N_{\rm HI}$ is treated
always as a free parameter minimizing $\chi^2$.)  We found that the
$\chi^2$ is minimized at $x_{\rm HI} = 0$ in all cases, favoring smaller
values.

\subsection{Spectral Index of the Afterglow SED}
\begin{figure}
  \begin{center}
    \FigureFile(80mm,80mm){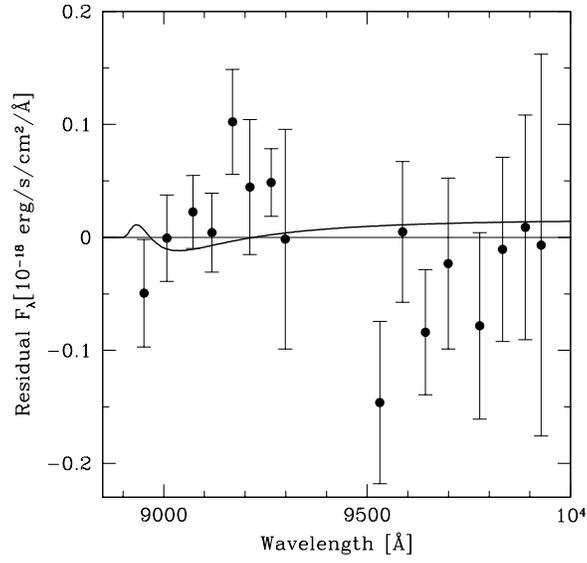}
  \end{center}
  \caption{The model and observed spectra of GRB 050904, in the form of
  the residual from the DLA-dominated model [($z_{\rm DLA}, \log N_{\rm
  HI}, z_{\rm IGM, u}, x_{\rm HI}$) = (6.295, 21.62, 6.295, 0)].  The
  connected pixels are binned by up to 20 pixels, after removing the
  wavelength ranges which were not used in the $\chi^2$ analyses because
  of absorption features. The thick solid curve is the best-fit model
  ($\log N_{\rm HI} = 21.54$) under the constraints of $x_{\rm HI} = 1$
  and $z_{\rm DLA} = z_{\rm IGM,u} = 6.295$.  }
  \label{fig:spec-residual}
\end{figure}

The change of the afterglow spectral index, $\beta_0$, within the quoted
error of $\Delta \beta_{\rm obs} = \pm 0.25$ (Tagliaferri et al. 2005)
does not have a significant effect.  As can be seen in
Fig. \ref{fig:spec-residual}, the data favors a bluer spectrum than the
baseline model with $\beta_0 = -1.25$, and hence a larger $\beta_0$
gives a better fit. However, this effect is not large; the change of
$\Delta \beta_0 = 0.25$ in the wavelength range of 9000--10000\AA \
results in the fractional change of $\sim 2.4$\% in $F_\lambda$, or
$\sim 0.024$ in Fig.  \ref{fig:spec-residual}.  It should be noted that
the quoted error of Haislip et al. (2005) is even smaller ($\Delta
\beta_{\rm obs} = ^{+0.15}_{-0.14}$). If the cooling frequency break in
the SED passed through the NIR bands at an epoch between the photometric
observations and our spectroscopic observation, the afterglow SED would
become even redder, which is not favored from our data.

\subsection{Dust Extinction in the Host Galaxy}

Though the observed NIR colors do not show evidence for extinction by
dust in the host galaxy (Tagliaferri et al. 2005), the wavelength
coverage by photometric observations redward of the Lyman break is
rather narrow, and substantial extinction may still be
possible. Therefore we tried a model with substantial extinction, where
the original afterglow spectrum $\beta_0$ is chosen so that the expected
NIR colors are consistent with the observed ones. The change of observed
$(J-K)$ color at $z \sim 6.3$ can be related to the restframe $V$
magnitude extinction as $\Delta (J-K) = 0.68 A_V$ and $1.83 A_V$ for the
extinction curves of the Milky Way (MW) and the SMC, respectively (e.g.,
Pei 1992). On the other hand, the change of the afterglow spectral index
is related to the NIR color as $\Delta (J-K) = - 0.61 \Delta \beta_0$.
Therefore the expected $(J-K)$ color is not changed if we introduce
extinction with changing $\beta_0$ by a relation, $A_V = 0.90 \Delta
\beta_0$ (MW) or $0.33 \Delta \beta_0$ (SMC). In Table \ref{table:chi2},
we show the fit results by models with $A_V = 0.45$ (MW) and 0.17 (SMC)
corresponding to $\Delta \beta_0 = +0.5$ ($\beta_0 = -0.75$).  Here we
have used the functional form of extinction $A(\lambda)$ around $\lambda
\sim \lambda_\alpha$ given in Barkana \& Loeb (2004).  We found that
both $\chi^2$ and $\Delta \chi^2$ are increased by introducing
extinction by dust, indicating that substantial extinction is not
favored and does not change the preference to lower $x_{\rm HI}$ values.

\subsection{The Redshift Parameters}

Changing $z_{\rm IGM,u}$ within the possible range of 6.27--6.314 does
not change the preference to lower $x_{\rm HI}$ values.  Because $z_{\rm
IGM,l}$ is rather uncertain, changing this parameter has a relatively
large effect. Larger values of $z_{\rm IGM,l}$ result in a smaller
amount of the neutral IGM and hence the upper limit on $x_{\rm HI}$
becomes weaker.  However, the completely neutral IGM can still be
excluded by $\sim$90\% C.L. even for $z_{\rm IGM,l} = 6.2$.

To show the characteristic redshift range of IGM neutral hydrogen
contributing to $\tau_{\rm IGM}$, we plot $\tau_{\rm IGM}$ as a function
of $z_{\rm IGM,l}$ with $z_{\rm IGM,u} =6.295$, for three values of
$\lambda_{\rm obs} = 8925$ (the lower bound of the wavelength range used
in the fitting analysis), 8970, and 9050 \AA, in
Fig. \ref{fig:tau_IGM_z_IGMl}.  The IGM optical depth is shown as a
ratio to that with $z_{\rm IGM, l} = 0$, which is $\tau_{\rm IGM} = $
0.37, 0.20, and 0.10 for the three values of $\lambda_{\rm obs}$,
respectively. From this figure we see that more than half and 80\% of the
IGM optical depth is contributed from hydrogens at $z \gtrsim 6.2$ and
6.0, respectively.

As a check for the possibility that $z_{\rm DLA}$ is different from
$z_{\rm metal}$ but $z_{\rm DLA} = z_{\rm host}$, we tested the models
with $z_{\rm DLA} = z_{\rm IGM, u} = 6.29$ and 6.314. This range can be
derived assuming $z_{\rm host} = z_{\rm IGM, u}$ and taking into account
the possible ranges of $z_{\rm host}$ and $z_{\rm IGM,u}$ with respect
to $z_{\rm metal}$, as well as the constraint from the transmission
around the Ly$\beta$ wavelength. We found that the preference to low
$x_{\rm HI}$ is not changed.

\subsection{Weak Unidentified Absorption Lines}

It is difficult to quantitatively estimate the possible systematic
effects by weak absorption lines that cannot be discriminated from
noise.  However, if such weak lines are distributed uniformly in the
fitting wavelength range, the effect is simply changing the overall
normalization that would not affect our conclusions. We also repeated
the same calculation without removing the wavelength ranges of the
identified absorption feature, to see the sensitivity to the absorption
features (see the ``Lines Included'' row of Table \ref{table:chi2}).
The data points are now 379, and $\chi^2$ is unacceptably large due to
the absorption features that are not included in the model. However, we
find a similar $\Delta \chi^2(1)$, and hence the preference to small
$x_{\rm HI}$ values is not sensitive to the treatment of absorption
lines.

The absorption by vibrationally excited molecular hydrogen may affect
the spectrum around the restframe Ly$\alpha$ (Draine 2000; Draine \& Hao
2002). In fact, Haislip et al. (2006)\footnote{In their e-print version
(astro-ph/0509660).} proposed that the anomalously attenuated $Z$-band
flux by a factor of $\sim 3$ of the early afterglow of GRB 050904 is due
to the molecular hydrogen absorption. However, this signature completely
disappeared by $\sim 3$ days, and hence it is unlikely to affect our
spectrum taken 3.4 days after the burst.  The molecular hydrogen
signatures have not yet been clearly detected in other afterglow spectra
(e.g., Schaefer et al. 2003; Vreeswijk et al. 2004).

\subsection{Time Variability}

Ionization by strong afterglow flux may result in the time variability
of the amount of hydrogen and metal absorptions (Perna \& Loeb 1998;
Draine \& Hao 2002). Though statistically significant variability was
not found in the observed spectrum of GRB 050904, variability within the
errors cannot be excluded. If such a variability of $N_{\rm HI}$ is
present, the observed shape of the damping wing is a superposition of
the shapes with different values of $N_{\rm HI}$. It may induce
systematic bias in the analysis presented here, since the damping wing
shape depends non-linearly on $N_{\rm HI}$.  As a test of this
possibility, we tried an extreme model where the spectrum is a mean of
the two spectra with different values of the column density, $N_+$ and
$N_-$, as $\log N_\pm \equiv \log N_{\rm HI} \pm \Delta \log N_{\rm
HI}$. We take $\Delta \log N_{\rm HI} = 0.2$ as the variability allowed
within the observational error, which is inferred from the statistical
error of $\log N_{\rm HI}$ obtained by the fitting analysis (see
Fig. \ref{fig:separate-contour}). We then find that the preference to
lower $x_{\rm HI}$ is hardly changed even for this rather extreme model
(see the ``Variability Check'' row of Table \ref{table:chi2}).

It should be noted that the neutral hydrogen shell must be very close to
the burster, as $r \lesssim 4 \times 10^{17}$ cm, in order for the
column density of $N_{\rm HI} \sim 21.6$ to substantially change during
the spectroscopic observation by the ionizing flux estimated from the
observed $F_\lambda$. This scale is close to the location of the
external shock at 3.4 days in observer's time for typical afterglow
parameters and $z = 6.3$ (e.g., Sari, Piran, \& Narayan 1998). We
derived in Paper I an estimate of electron density $n_e \sim 10^{2.3 \pm
0.7} \ \rm cm^{-3}$ from the observed fine-structure Si* line, assuming
collisional excitation by electrons. The ionization fraction is unknown,
and if ionization fraction is much smaller as $n_e/n_H \lesssim
10^{-4}$, the excitation by neutral hydrogen becomes more important,
leading to a density estimate of $n_H \sim 10^4 \ \rm cm^{-3}$ (Silva \&
Viegas 2002). Combined with the DLA column density, the DLA scale of
$\sim$ 0.1--1 pc is inferred. On the other hand, Berger et al. (2005b)
suggested a possibility of fine-structure excitation by GRB afterglow
radiation for the case of GRB 051111, deriving a larger distance scale
of $\sim$10--20 pc. Therefore the scale estimate is rather uncertain,
but it seems that there is a parameter space of the DLA where no $N_{\rm
HI}$ variability is expected but it is close enough to be consistent
with the Si$\emissiontype{II}$* detection.

Based on these results, we conclude that there is an evidence for
considerable ionization of the IGM around the host galaxy of GRB 050904
already at $z=6.3$.

\section{Discussion}
\label{section:discussion}

\subsection{Prediction for the Case of Low $N_{\rm HI}$}
\label{section:discussion-DLA}

We found that the DLA column density is $\log N_{\rm HI} \gtrsim 21$
for GRB 050904, and
in such a case the DLA absorption dominates the absorption by neutral
IGM extending to the same redshift as that of the DLA. Therefore it is
difficult to derive a very strong constraint on $x_{\rm HI}$, 
though we obtained some constraints as presented above by
detailed statistical analyses.  Vreeswijk et al. (2004) compiled seven
GRBs having estimates of $N_{\rm HI}$ from afterglow spectra, and though
the majority of them have column densities larger than $\log N_{\rm HI}
\geq 21$, two (GRB 011211 and GRB 021004) have low column densities of
$\log N_{\rm HI} \lesssim 20$. If we detect such a GRB at $z \gtrsim 6$,
it would provide an opportunity to probe the reionization with a much
cleaner environment.

\begin{figure}
  \begin{center}
    \FigureFile(90mm,90mm){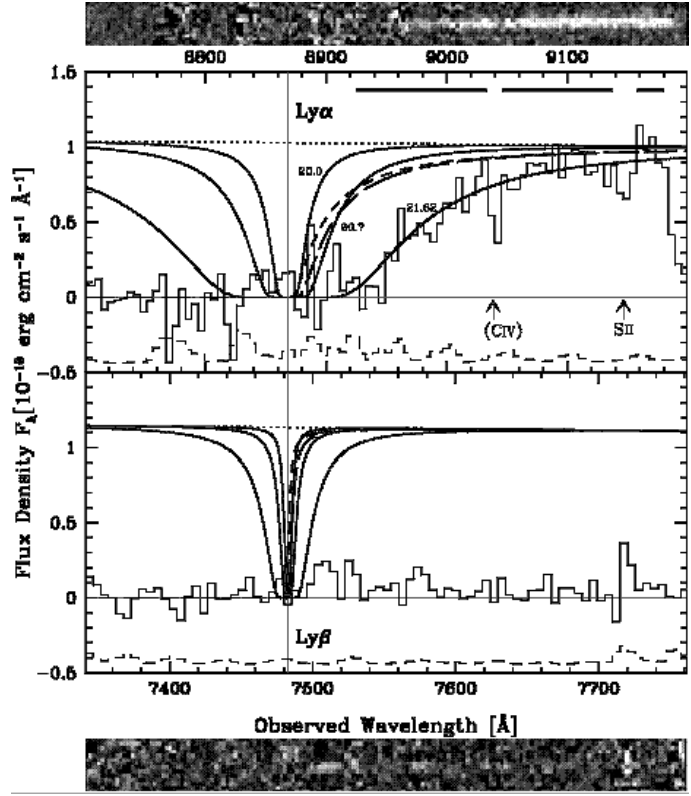}
  \end{center}
  \caption{The same as Fig. \ref{fig:spec2}, but showing some 
  different models. The three solid curves are the model absorption
  by the DLA at $z_{\rm DLA} = 6.295$, but with different values of
  $\log N_{\rm HI} = 20.0, 20.7, $ and 21.62, 
   as indicated. The short-dashed curve shows the
   model absorption by the IGM with $z_{\rm IGM,u} = 6.295$ and
   $x_{\rm HI} = 1.0$, and the long-dashed curve is the same but
   the DLA absorption with $\log N_{\rm HI} = 20$
   and $z_{\rm DLA} = 6.295$ is added. 
  } \label{fig:spec3}
\end{figure}

To demonstrate this possibility, we show a model with $(\log N_{\rm HI},
x_{\rm HI}) = (20.0, 1.0)$ with $z_{\rm DLA} = z_{\rm IGM,u} = z_{\rm
metal} = 6.295$, in Fig. \ref{fig:spec3}. In this case the shape of the
damping wing is dominantly contributed by the neutral IGM
absorption. Also shown is DLA absorption at the same redshift but with
various values of $N_{\rm HI}$. It can be seen that the shape of IGM
absorption is different from that of DLAs with any values of $N_{\rm
HI}$, making it possible to prove the dominance of the IGM
component. Another possibility to explain such a data by DLA absorption
is to consider a blueshifted absorption by a DLA with a relative
velocity of a few thousands km/s with respect to $z_{\rm metal}$, as we
observed the degeneracy between the DLA and IGM absorptions in \S
\ref{section:separate-fit}.  Such a velocity may again be achieved by
acceleration of absorbing clouds by the activity of GRBs or their
progenitors, but such absorbers are likely to be polluted by metals, and
we expect blueshifted metal lines as well, in addition to those in the
restframe of the host galaxy.  Therefore we can discriminate this case
as well, and hence high-quality spectra of low-$N_{\rm HI}$ afterglows
in the future will provide us with an invaluable opportunity to
measure $x_{\rm HI}$ more accurately.

\subsection{Ly$\alpha$ Emission from Host Galaxies and Reionization}

\label{section:discussion-lya-emission}

We expect Ly$\alpha$ emission in the spectrum, if the huge
H\emissiontype{I} absorption is by gas in the vicinity of the GRB and
the host galaxy has detectable Ly$\alpha$ emission like GRB 030323
(Vreeswijk et al. 2004).  The spectrum of GRB 050904 does not show a
clear Ly$\alpha$ emission (\S \ref{section:spectrum}) and we set an
upper limit on the Ly$\alpha$ flux of $F(\rm Ly\alpha) < 1.7 \times
10^{-18} \ \rm erg \ cm^{-2} s^{-1}$ for a width of 8.5 \AA, which is
the FWHM of the spectral resolution and corresponding to a velocity
dispersion of $\sim$ 300 km/s as well. Then the upper bound on the
Ly$\alpha$ luminosity becomes $L(\rm Ly\alpha) < 7.9 \times 10^{41}$
erg/s at $z = 6.295$, which can be translated into
extinction-uncorrected SFR of $< 0.79 M_\odot$/yr, by using the
empirical relation (Kennicutt 1998; Cowie \& Hu 1998).

It has been indicated that the Ly$\alpha$ emission from GRB host
galaxies is stronger in terms of the equivalent width (EW) than that
from Lyman-break galaxies (LBGs) at similar redshift ($z \sim 3$); all
GRB host galaxies at $z \gtrsim 2$ are consistent with having EW$_{\rm
rest} ({\rm Ly}\alpha) > 10$ \AA \ (Fynbo et al. 2003; Jakobsson et
al. 2005), which is as large as that of LAEs found in deep surveys.  For
comparison, only about a third of LBGs at similar redshifts have such
large EW (Shapley et al. 2003). Then it is suggested that Ly$\alpha$
emission from host galaxies of GRBs at $z \gtrsim 6$ may also give some
information on the reionization.

To demonstrate this possibility, we show the IGM absorption profile with
$z_{\rm IGM,u} = 6.295$ for $x_{\rm HI} = 0.01, 0.2$, and 1.0 in
Fig. \ref{fig:spec2}.  If $x_{\rm HI} \sim 1$ and the Ly$\alpha$ line
center is the same as $\lambda_\alpha (1 + z_{\rm IGM,u})$, the
Ly$\alpha$ photons within $\sim$10 \AA \ from the line center are
heavily absorbed. For comparison, LAEs detected at $z \sim 6$ show line
widths of $\sim$ 10\AA, corresponding to a few hundreds km/s (Kodaira et
al.  2003; Taniguchi et al. 2005). Even larger velocity dispersion may
also be possible; a redward tail of Ly$\alpha$ emission extending beyond
a relative velocity difference of $\gtrsim $ 1000 km/s has been found in
a starburst galaxy at $z \sim 3$ (Pettini et al. 2000). Then we expect
that, if $x_{\rm HI} \sim 1$, the Ly$\alpha$ line luminosity is strongly
attenuated allowing transmission only in the redward tail. On the other
hand, if $x_{\rm HI} \lesssim 0.01$, about half of the Ly$\alpha$
emission is transmitted. Therefore statistical comparison of Ly$\alpha$
line emissivity of GRB host galaxies at 
$z \gtrsim 6$ and $\lesssim 6$  may
give an interesting information for the reionization.

As a reionization probe based on Ly$\alpha$ emission, GRB host galaxies
have an advantage of no selection effect compared with LAEs selected by
Ly$\alpha$ emission. Brightness of afterglows allows us to accurately
measure the redshift by metal absorption lines, as demonstrated by GRB
050904, and hence the plausible Ly$\alpha$ line center can be determined
more accurately than LAEs or LBGs.  It should be noted that, even though
LAEs at $z \sim 6$ have asymmetric Ly$\alpha$ emission profiles with a
sharp cut-off at the blueward side, a strong conclusion about the IGM
neutral fraction cannot be derived because of the general lack of
information about the line center (Haiman 2002). On the other hand,
since the brightness of host galaxies is not relevant to the
detectability of GRBs, typical GRBs may occur in smaller galaxies than
LAEs/LBGs, and hence typical absolute Ly$\alpha$ luminosity may be
smaller, in spite of the suggestion of large EW for GRB host
galaxies. (See \S\ref{section:z_IGMu_z_metal} for a discussion about the
typical mass of GRB host galaxies at $z \sim 6$ expected from the
structure formation theory.)  A larger sample of GRB host galaxies at $z
\gtrsim 6$ is necessary to investigate more about these possibilities.

Concerning the case of GRB 050904, a search for the host galaxy in the $z'$
band will be interesting in this context. Assuming $\rm EW_{\rm rest}
(Ly\alpha) \geq 10$ \AA \ as suggested for GRB host galaxies, the obtained
upper bound on Ly$\alpha$ emission requires that the continuum level of
the host galaxy must be lower than $F_\lambda \leq 2.3 \times 10^{-20}
({\rm EW_{rest}}/{\rm 10\AA})^{-1} \ \rm erg \ cm^{-2} s^{-1}
\AA^{-1}$. Since the Lyman break is in the midst of the $z'$ band filter, we
take into account a dimming by a factor of about 2, and we get the
corresponding AB magnitude of $z' > 27.7$.  Therefore, if follow-up
observations find the host galaxy of GRB 050904 brighter than $z' =
27.7$, it would mean that $\rm EW_{\rm rest}(Ly\alpha) < 10 \AA$, which
is smaller than those found in GRB host galaxies at $z \sim 3$. Thus
it indicates a possibility of attenuation by neutral IGM, though
only one case is not sufficient and 
statistical studies are required. Galaxies
brighter than $z' = 27.7$ can be detected by imaging observations of
existing large telescopes; typical magnitude of LAEs at $z \sim 6.6$
found in the Subaru Deep Field is $z' \sim$ 26--27.8 (Taniguchi et
al. 2005).

\section{Summary and Conclusions}
\label{section:summary}

We presented a comprehensive theoretical modeling of the red damping
wing of the Ly$\alpha$ absorption found in the optical afterglow
spectrum of GRB 050904 at $z \sim 6.3$, which provides the first
opportunity of studying the cosmic reionization by using GRBs.  We tried
to model the observed damping wing shape by the two components of
absorbers: one is by the DLA associated to the GRB host galaxy, and the
other is by neutral hydrogen in the IGM. The redshift of the metal
absorption lines in the spectrum is $z_{\rm metal} = 6.295 \pm 0.002$,
but we allowed different values of $z_{\rm DLA}$ (the DLA redshift) and
$z_{\rm IGM,u}$ (the upper extension bound of neutral hydrogens in the
IGM), and discussed various theoretical possibilities for the deviation
of these parameters from $z_{\rm metal}$.

The shape of the red damping wing can be explained either by the DLA at
$z_{\rm DLA} \sim z_{\rm metal}$ with $\log N_{\rm HI} \sim $ 21.6, or
by almost neutral IGM extending to a higher redshift of $z_{\rm IGM,u}
\sim $ 6.36. Though the DLA seems a more straightforward interpretation,
we cannot exclude the latter possibility simply by the redshift
difference, since blueshift of metal absorption lines up to a few
thousands km/s with respect to the restframe of the host galaxy has been
observed in a few GRBs.

However, we found that the Ly$\beta$ feature can be used to break this
degeneracy, since the two different solutions predict different
wavelengths at which the Ly$\beta$ GP trough ends.  Then we
concluded that the damping wing is mostly contributed from the DLA at
$z_{\rm DLA} \sim z_{\rm metal}$, and derived a firm upper bound of
$z_{\rm IGM,u} \leq 6.314$.  We argued that the DLA is likely to be
associated physically with the metal absorption lines, and the inferred
column densities, metallicities, and depletion of silicon are all
reasonable as a DLA found in a GRB afterglow.

Next we examined the preferred value of the IGM neutral fraction,
$x_{\rm HI}$, in the viable model of $z_{\rm IGM,u} = z_{\rm DLA} =
6.295$. Treating $N_{\rm HI}$ of the DLA as a free parameter, we found
that a smaller value of $x_{\rm HI}$ is favored with the best-fit value
of $x_{\rm HI} = 0.00$, and upper limits of $x_{\rm HI} < 0.17$
and 0.60 (68 and 95\% C.L., respectively) were derived.  We examined
various possible systematic uncertainties that could affect this result,
including the afterglow spectral index, dust extinction at the host
galaxy, the redshift parameters of the DLA and IGM absorptions, weak
unidentified absorption lines, and time variability of the DLA column
density. We found that none of these effects is large enough to change
the above result. Hence we conclude that the universe was largely
ionized already at $z \sim 6.3$, excluding the completely neutral IGM at
$\sim$ 99\% C.L.

\begin{figure}
  \begin{center}
    \FigureFile(80mm,80mm){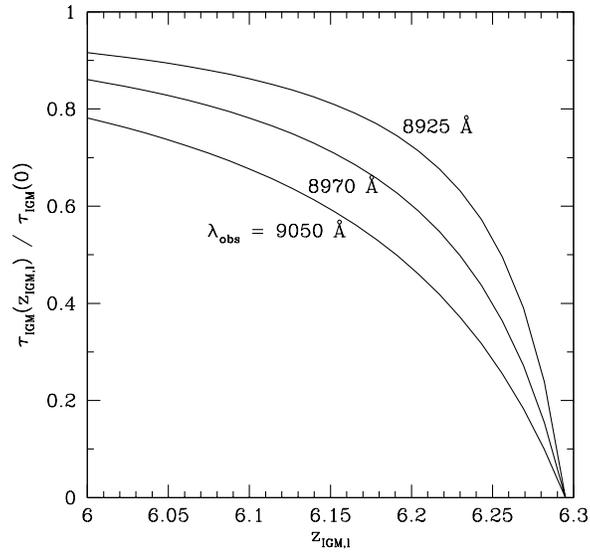}
  \end{center}
  \caption{The IGM optical depth as a function of $z_{\rm IGM,l}$, which
  is normalized by that at $z_{\rm IGM,l} = 0$, is shown for three
  values of $\lambda_{\rm obs}$ as indicated. Here we set $z_{\rm IGM,u}
  = 6.295$.  } \label{fig:tau_IGM_z_IGMl}
\end{figure}

This is the first quantitative {\it upper} limit on $x_{\rm HI}$ at
$z \gtrsim 6$ by a direct method,\footnote{Though there are some
``data points'' of the IGM optical depth at $z \gtrsim 6$
derived directly from
the GP trough of quasar spectra (Fan et al. 2006), these are by averaging
sharp spikes of transmission in a wavelength range,
which are probably corresponding to small regions
of ionized bubbles. Therefore there is no upper limit on the 
mass-weighted or volume-weighted optical depth.} 
being consistent with the recent
results by an indirect approach using the number density evolution of
LAEs (Malhotra \& Rhoads 2004; Stern et al. 2005; Haiman \& Cen 2005).
Since the IGM is optically thin for photons in the damping wing region,
all the IGM neutral hydrogens at $z \sim 6.1$--6.3 (see
Fig. \ref{fig:tau_IGM_z_IGMl}) contribute to the damping wing, allowing
us to derive a robust constraint on the mass-weighted $x_{\rm HI}$ which
is insensitive to any clumpiness of IGM within this redshift interval.
Combined with the suggestions of $x_{\rm HI} \gtrsim 0.1$ from quasar
spectra (Mesinger \& Haiman 2004; Wyithe, Loeb, \& Carilli 2005), a
plausible value of $x_{\rm HI} \sim 0.1$ is suggested for the IGM at $z
\sim 6$--6.3.

The large DLA column density of $\log N_{\rm HI} \gtrsim 21$ dominates
the IGM absorption extending to the same redshift, making it difficult
to derive a stronger constraint on $x_{\rm HI}$ than derived
here. However, some GRBs have low column densities of $\log N_{\rm HI}
\lesssim 20$, and detection of such GRBs at $z \gtrsim 6$ will be a
promising chance to get better information for the reionization history
of the universe.

We did not detect Ly$\alpha$ emission from the host galaxy, leading to
an upper limit for extinction-uncorrected star formation rate as SFR
$\lesssim 0.79 M_\odot$/yr.  We discussed the potential of Ly$\alpha$
emission from GRB host galaxies as a reionization probe.  Statistically
smaller equivalent width and transmission only at the red tail of
Ly$\alpha$ emission are expected for GRB host galaxies before the
reionization, compared with those at lower redshifts. This may be tested
by using future large samples of GRBs at $z \gtrsim 6$.  As a
reionization probe using Ly$\alpha$ emission, GRB host galaxies have an
advantage of being free from the selection bias compared with
LAEs. Another advantage compared with both LAEs and LBGs is that an
accurate redshift determination is possible by absorption lines in
bright afterglows. On the other hand, a possible disadvantage is that
typical GRB host galaxies and their absolute
Ly$\alpha$ luminosity may not be as bright as LAEs and LBGs found in
deep surveys.

We conclude that the GRB 050904 has opened a new era of cosmological
study by GRBs, and future data will give us even more unique and important 
information about the epoch when the early-generation luminous
objects changed the physical state of almost all the baryonic matter
in the cosmos.

We would like to thank the Subaru Telescope staff for their warm
assistance in taking this invaluable data. We would also like to thank the
referee for useful comments. This work was supported in part by the
Grant-in-Aid for the 21st Century COE ``Center for Diversity and
Universality in Physics'' from the Ministry of Education, Culture,
Sports, Science, and Technology (MEXT) of Japan.  T.T. and N.K. were also
supported by the Grant-in-Aid for Scientific Research from the MEXT,
16740109 and 14GS0211, respectively.


\onecolumn

\begin{table}
  \small
  \begin{center}
   \caption{Constraint on $x_{\rm HI}$ for Various Models
      \label{table:chi2}
}
   \begin{tabular}{lccccccc}
     \hline \hline
     & & & & & \multicolumn{3}{c}{upper limits
           on $x_{\rm HI}$\footnotemark[$*$]}   \\
     \cline{6-8} 
      Models & $\chi^2_{\min}$ \footnotemark[$\dagger$]  
         & $\chi^2(x_{\rm HI}=1)$  
         & $\Delta \chi^2(x_{\rm HI}=1)$
         & C.L.(\%)\footnotemark[$\ddagger$]  &  68 & 95 & 99 C.L. (\%) \\
     \hline Baseline\footnotemark[$\S$] &  277.29 & 284.12 &  6.84 & 99.1
         & 0.17 & 0.60 & 0.98 \\
     $\beta_0 = -1/-1.5$ &  273.86/281.10 &  280.07/288.75 & 6.21/7.65 
          & 98.7/99.4 & 0.18/0.15 & 0.66/0.52 & 1.08/0.88  \\
     $A_V = 0.45/0.17$\footnotemark[$\P$] 
         & 316.93/283.18 & 329.72/291.12 &  12.80/7.94
         & 99.97/99.5 &  0.09/0.15 & 0.32/0.52 & 0.54/0.85 \\
     $z_{\rm IGM,u} = 6.27/6.314$  & 277.29/277.29 & 283.55/283.21 
           &  6.27/5.93 & 98.8/98.5  & 0.19/0.19 & 0.63/0.68 & 1.06/1.12 \\
     $z_{\rm IGM,l} = 5.5/6.2$ & 277.29/277.29 & 286.70/280.07 & 9.41/2.78
          & 99.8/90.5 & 0.13/0.38 & 0.45/1.36 & 0.74/2.22 \\
     $z_{\rm DLA} = 6.29/6.314$\footnotemark[$\Vert$]  
          & 281.47/266.41 & 288.49/271.25 
          & 7.03/4.83 & 99.2/97.2 & 0.16/0.27 & 0.57/0.84 & 0.96/1.27 \\
     Lines Included & 796.03 & 802.53 & 6.50 & 98.9 & 0.17 & 0.63 & 1.02 \\
     Variability Check & 283.98 & 289.54 & 5.56 & 98.2 & 0.20 & 0.71 & 1.18 \\
     \hline
     \hline
     \multicolumn{8}{@{}l@{}}{\hbox to 0pt{\parbox{180mm}{\footnotesize

    \vspace{0.2cm}

    \par\noindent
    \footnotemark[$*$]See \S\ref{section:model} for the reason why the
         apparently unphysical cases of $x_{\rm HI} > 1$ are allowed here.

    \par\noindent
    \footnotemark[$^\dagger$]In all models, $\chi^2$ increases with
       $x_{\rm HI}$ and the minimum $\chi^2$ is realized at $x_{\rm HI} =0$.

    \par\noindent
    \footnotemark[$\ddagger$]The exclusion confidence level for
    the case of $x_{\rm HI} = 1$. 

    \par\noindent
    \footnotemark[$\S$]The baseline model parameters are: 
       $z_{\rm IGM, u} = z_{\rm DLA} = 6.295$, 
       $z_{\rm IGM, l} = 6$, $\beta_0 = -1.25$, and $A_V = 0$. 
       The second row and below show the models
       when some parameters are changed from the baseline model.

    \par\indent
    \footnotemark[$\P$]The spectral index is changed into 
       $\beta_0 = -0.75$ to keep
       the expected NIR colors consistent with the observed ones
       for the MW/SMC extinction curves, respectively.

    \par\indent
    \footnotemark[$\Vert$]The IGM redshift parameter $z_{\rm IGM, u}$
       is kept to be the same with $z_{\rm DLA}$.

    }\hss}}
   \end{tabular}
  \end{center}
\end{table}


\end{document}